\PassOptionsToPackage{unicode}{hyperref}
\PassOptionsToPackage{hyphens}{url}
\PassOptionsToPackage{dvipsnames,svgnames,x11names,table}{xcolor}

\documentclass[9pt,arxiv]{lapreprint}

\usepackage[version=4]{mhchem} 
\usepackage{siunitx}    
\usepackage{pdflscape}  
\usepackage{rotating}   
\usepackage{textgreek}  
\usepackage{gensymb}    
\usepackage[misc]{ifsym} 
\usepackage{listings}   
\usepackage{colortbl}   
\usepackage{tabularx}   
\usepackage{subcaption}
\usepackage{multirow}
\usepackage{snotez}     
\usepackage{natbib}

\usepackage[sfdefault,condensed]{roboto}
\DeclareSIUnit\Molar{M}

\usepackage{hyperref} 
\usepackage{graphicx}

\makeatletter
\def\maxwidth{\ifdim\Gin@nat@width>\linewidth\linewidth\else\Gin@nat@width\fi}
\def\maxheight{\ifdim\Gin@nat@height>\textheight\textheight\else\Gin@nat@height\fi}
\makeatother
\setkeys{Gin}{width=\maxwidth,height=\maxheight,keepaspectratio}
\makeatletter
\def\fps@figure{htbp}
\makeatother

\usepackage{etoolbox}
\BeforeBeginEnvironment{longtable}{\begin{center}\scriptsize}
\AfterEndEnvironment{longtable}{\end{center}}

\newlength{\cslhangindent}
\newlength{\csllabelwidth}
 {
  \setlength{\parindent}{0pt}
  \ifodd #1 \everypar{\setlength{\hangindent}{\cslhangindent}}\ignorespaces\fi
  \ifnum #2 > 0
  \setlength{\parskip}{#2\baselineskip}
  \fi
 }%
 {}
\usepackage{calc}

\usepackage{longtable}
\usepackage{booktabs}
\usepackage{tabularx}
\usepackage{calc}

\title{Reducing Urban Speed Limits Decreases Work-Related Traffic Injury Severity: Evidence from Santiago, Chile}

\author[1,2]{Eduardo Graells-Garrido}

\author[1]{Mat\'ias Toro}

\author[3]{Gabriel Mansilla}
\author[3]{Mat\'ias Nicolai}
\author[3]{Santiago Mansilla}

\author[4]{Jocelyn Dunstan}

\affil[1]{Department of Computer Science, University of Chile, Santiago, Chile}
\affil[2]{National Center for Artificial Intelligence (CENIA), Santiago, Chile}
\affil[3]{GSE Salud Consultores, Santiago, Chile}
\affil[4]{Department of Computer Science \& Institute for Mathematical Computing, Pontificia Universidad Católica de Chile, Santiago, Chile}

\usepackage{libertinus}

\metadata[]{\Letter\hspace{.5ex} For correspondence}{egraells@dcc.uchile.cl}

\metadata[]{Acknowledgments}{This work was selected in the Call for Research and Innovation Projects
in Accident Prevention and Occupational Diseases (2022) by the
Superintendency of Social Security of Chile (SUSESO), and was funded by
Mutual de Seguridad (CChC) with resources from the Social Security of
the Law No. 16,744 on Occupational Accidents and Diseases. E. Graells-Garrido was partially funded by National Center for Artificial Intelligence CENIA FB210017, Basal ANID. The authors thank the administrative and technical management from Luis Alarc\'on at Mutual de Seguridad.}

\metadata[]{Competing interests}{The author(s) declared no potential conflicts of interest with respect
to the research, authorship, and/or publication of this article.}

\leadauthor{Graells-Garrido}
\shorttitle{Reducing Urban Speed Limits Decreases Work-Related Traffic Injury Severity}

\begin{document}
\maketitle

\begin{abstract}
  Work-related transportation incidents significantly impact urban mobility and productivity. These incidents include traffic crashes, collisions between vehicles, and falls that occurred during commuting or work-related transportation (e.g., falling while getting off a bus during the morning commute or while riding a bicycle for work). This study analyzes a decade of work-related transportation incident data (2012--2021) in Santiago, Chile, using records from a major worker's insurance company. Using negative binomial regression, we assess the impact of a 2018 urban speed limit reduction law on incident injury severity. We also explore broader temporal, spatial, and demographic patterns in these incidents in urban and rural areas.

  The urban speed limit reduction is associated with a decrease of 4.26 days in prescribed medical leave for incidents in urban areas, suggesting that lower speed limits contribute to reduced injury severity. Our broader analysis reveals distinct incident patterns across different groups. Workers traveling by motorcycle and bicycle experience more severe injuries when involved in traffic incidents, with marginal effects of 26.94 and 13.06 additional days of medical leave, respectively, compared to motorized vehicles. Women workers tend to have less severe injuries, with an average of 7.57 fewer days of medical leave. Age is also a significant factor, with older workers experiencing more severe injuries --- each additional year of age is associated with 0.57 more days of medical leave. Our results provide insights for urban planning, transportation policy, and workplace safety initiatives.
\end{abstract}

Work trips are a crucial aspect of daily urban mobility, shaping the rhythm of cities and defining peak travel periods. The safety of these trips is important, as work-related {traffic incidents (including crashes between vehicles, collisions with other road users, and falls during commuting or work transportation)} impact individual health, urban productivity, and sustainability \citep{unGoalDepartment}. These incidents significantly impact urban dynamics by affecting workers, disrupting traffic, straining emergency services, and causing economic losses through reduced productivity \citep{chantith2021measure}. In this paper, we use the term `incident' to refer to traffic incidents that may result in injury or property damage, avoiding the term `accident' which can imply unavoidability.

While disruptive to urban systems, traffic incidents are often underrepresented in transportation analyses. Traditional travel surveys rarely capture traffic incident data, and new data sources like mobile phone records, although helpful in understanding general mobility patterns \citep{widhalm2015discovering,graells2021city,graells2023data}, cannot identify specific traffic incidents due to anonymization and lack of meta-data. However, work-related incidents offer study opportunities, as they are more likely to be centrally recorded due to legal and insurance requirements.

Our study focuses on Santiago, the capital of Chile and its largest urban agglomeration. According to the last official Census from 2017, the official population of Santiago is of nearly 8 million inhabitants spread across an urban area of 867.75 square kilometers, comprising over 40 independent municipalities. The city's transportation system is characterized by an integrated multimodal network with a nearly flat fare structure across the metro, urban buses, and rail services \citep{munoz2014transantiago}.

Various institutions prioritize worker safety and insurance in Santiago, overseen by the \emph{Superintendencia de Seguridad Social} (SUSESO). This paper analyzes a dataset from one of these entities, \emph{Mutual de Seguridad de la Cámara Chilena de la Construcción} (MUSEG), which covers approximately 31\% of all insured workers in the country as of 2022. {We quantify incident severity through prescribed medical days of leave --- the time a medical professional determines a worker needs to recover before returning to work. This measure directly captures the immediate impact on workers' lives and productivity}. {Our study examines 75.8K traffic incidents in the Santiago metropolitan area and surrounding rural regions (of which 33.4K could be geolocated with exact coordinates)} from 2012 to 2021, revealing temporal, geographical, and demographic patterns.

A significant event within our study period is the implementation of a new speed limit regulation on August 4th, 2018, which reduced the urban speed limit from 60 km/h to 50 km/h \citep{ReduccionVelocidad}. {While this law directly regulates motorized vehicle speeds, its safety benefits extend to all road users, as most crashes involving pedestrians, cyclists, and motorcyclists also involve cars or other motor vehicles on shared roads}. We use this policy change as a natural experiment, employing a Negative Binomial regression to assess its effect on incident severity.

This work addresses the following research questions:

\begin{enumerate}
    \item What are the temporal, spatial, and demographic patterns of work-related transportation incidents in Santiago?
    \item How did the 2018 speed limit reduction impact the severity of work-related transportation incidents?
    \item How do factors such as mode of transportation, time of day, and worker demographics influence incident severity?
\end{enumerate}

This paper is structured as follows: Section 2 reviews related work on urban mobility safety and speed limit impacts. Section 3 describes our context and data sources. Section 4 describes the methodology. Section 5 presents our results, including the spatial and temporal patterns of incidents and the regression analysis of the impact of the speed limit law. Section 6 concludes with key takeaways and suggestions for future research.

\section{Related Work}

{Research on urban transportation safety has shown that speed limit reductions can effectively reduce traffic injuries} \citep{elvik2012speed, cleland2020effects, milton2021use}. {However, less than 10\% of road safety research focuses on low- and middle-income countries, despite these areas accounting for most road traffic deaths and injuries} \citep{haghani2022road}. {Our study addresses this gap by examining how speed limit changes affect work-related transportation incidents in a Latin American context, while also investigating the broader relationship between urban mobility patterns and traffic safety}.

\subsection{Urban Mobility and Safety}
Urban mobility patterns significantly influence traffic safety. \citet{miner2024car} categorize the negative impacts of automobility into violence, ill health, social injustice, and environmental damage. Their study reveals that cars have contributed to 60–80 million deaths globally and exacerbated social and environmental issues. They advocate for replacing car-centric policies with interventions such as increasing bike lanes or reducing traffic speed limits.

\citet{cabrera2020uncovering} and \citet{cabrera2021urban} explored the relationship between urban size and incident rates. They found that while the number of incidents in an urban area generally depends superlinearly on population size, this relationship varies with incident severity. Notably, they found no significant change in the number of road traffic collisions per person for urban areas of different sizes, highlighting the complex nature of urban mobility and safety.

\subsection{Spatial Analysis of Traffic incidents}

Spatial patterns of traffic incidents provide crucial insights for targeted interventions. \citet{hazaymeh2022spatiotemporal} used local hotspot analysis to study traffic incidents in Jordan, concluding that less severe incidents occur where traffic volume is highest.

In the Latin American context, \citet{mangones2024safety} compared crash risk on arterial roads, Bus Rapid Transit (BRT), and Buses with High Level of Service (BHLS) corridors in Bogotá, Colombia. Their findings suggest that different public transport infrastructures have varying impacts on safety. BRT networks provide lower crash rates for less severe collisions but increase injuries and fatalities. This underscores the importance of considering different road types and transport modes in safety analyses. \citet{sukhai2021fatality}, studying South Africa, another Global South country, found that contextual effects create unequal road safety risks, with higher vulnerabilities for pedestrians, women, and public transport users linked to rurality and socio-economic factors.

\subsection{Impact of Speed Limit Laws}

{Research on speed limit reductions, particularly from the UK, provides strong evidence for their effectiveness. In Edinburgh, the introduction of 20 mph (32 km/h) limits resulted in a 22\% reduction in collisions and 20\% reduction in casualties compared to control zones} \citep{kokka2024effect}. {A comprehensive meta-narrative synthesis found that 20 mph zones were consistently associated with reductions in collisions and casualties} \citep{cleland2020effects}. {Multiple studies from Bristol and Edinburgh demonstrated that speed limit reductions led to measurable decreases in average traffic speeds and crashes} \citep{pilkington2018bristol, nightingale2020op90}.

{While previous studies have shown the effectiveness of speed limit reductions in various} contexts \citep{elvik2012speed, milton2021use, cleland2020effects}, {including rigorous evaluations of citywide implementations} \citep{kokka2024effect, pilkington2018bristol}, {research examining speed enforcement mechanisms provides additional supporting evidence}. \citet{montella2015effects} {found that point-to-point speed enforcement on urban motorways reduced excessive speeding behaviors by over 80\% and crashes by 32\%, with even larger reductions during adverse conditions. Similarly,} \citet{goldenbeld2005effects} {documented a 21\% reduction in both injury accidents and serious casualties after implementing targeted speed enforcement}.

In the Chilean context, \citet{martinez2020effects} and \citet{nazif2014explaining} studied the impacts of a 2005 traffic law reform. \citet{martinez2020effects} found that the reform led to significant reductions in road traffic death trends for both child pedestrians and passengers, with socioeconomic factors also playing a role. \citet{nazif2014explaining} observed that the 2005 reform was associated with a 7\% reduction in pedestrian fatalities, mediated by increased police enforcement and reduced alcohol consumption. They also found that police enforcement was directly associated with significant decreases in fatalities across all road user types, while road infrastructure investment was linked to an 11\% reduction in pedestrian fatalities. These studies provide essential context for understanding the evolving landscape of traffic safety regulations in Chile and highlight the complex interplay between legislative changes, enforcement, and infrastructure investment in improving road safety outcomes.

\subsection{Gender and Age Dynamics in Traffic Incidents}

\citet{gonzalez2021traffic} investigated the influence of road type on traffic injury risk, considering gender and age. They found that men consistently have a higher risk of severe and fatal injuries across all modes of transport and road types. \citet{havet2021gender} explored how determinants of daily mobility affect men and women workers differently, finding persistent gender differences in travel patterns even when controlling for employment status and car access. 

\citet{Salminen2000Traffic} found that in Finland, men were involved in five out of every six traffic incidents during work hours, despite driving only three out of every four kilometers. The study also revealed that older workers (50-65 years) had the highest frequency of work-related transportation incidents, highlighting the importance of age as a factor in our analysis.

\subsection{Work-Related Incidents in Urban Contexts}
Work-related transportation incidents, particularly those occurring during commutes, represent a significant subset of urban traffic incidents with unique characteristics and impacts. These incidents affect individual workers and have broader implications for urban mobility and economic productivity.

\citet{mitchell2004work} conducted a comprehensive study of work-related road deaths in Australia, finding that, on average, 151 commuters were fatally injured in vehicle incidents on public roadways each year from 1989 to 1992. This translates to a rate of 2.0 per 100,000 commuters per year, highlighting the significant risk associated with daily work-related travel in urban areas.

The economic impact of commuting incidents is substantial. While not specific to commuting, \citet{chantith2021measure} calculated that productivity loss due to road traffic incidents in Thailand amounted to approximately 0.8\% of the country's GDP in 2017. This underscores the significant economic burden that traffic incidents, including those related to commuting, can impose on urban economies.

The duration of medical leave following commuting incidents is a critical aspect affecting workers and employers. \citet{hansson1976sick} studied road traffic casualties in Sweden and found that the average duration of inability to work was 35 days. They noted a close relation between the seriousness of the injury and time off work, with orthopedic injuries resulting in an average of 120 days of medical leave. Although this study is not recent, it highlights the potentially long-term impact of traffic incidents on workers' ability to return to work.

Two years post-crash, 8\% of car occupants in Sweden {had permanent medical impairment} \citep{elrud2019sickness}. For cycle crashes, 20\% of individuals had sickness absence, with duration varying by injury type and body region \citep{ohlin2018duration}. These studies highlight the long-term impacts of traffic crashes on workers.

Urban infrastructure plays a crucial role in commuting safety. \citet{mangones2024safety} explored how different public transport infrastructures in Bogot\'a, Colombia, impact safety. They found that different bus corridor types affect crash rates and severity differently, highlighting the importance of infrastructure design in commuting safety.

The gender aspect of commuting safety is a critical consideration in urban mobility. \citet{gauvin2020gender} studied urban mobility patterns in Santiago, Chile, revealing a gender gap in mobility. Women were found to visit fewer unique locations than men and distribute their time less equally among such locations. This gender difference in mobility patterns could affect commuting safety. Complementing this, \citet{vasquez2020tweets} analyzed social media data from Santiago to understand gender differences in transportation experiences. They found that women more frequently express concerns about safety and harassment in public transportation and public spaces. This aligns with findings from \citet{loukaitou2016gendered}, who highlight that women's transport needs differ from men's due to safety concerns and sociocultural norms. These studies underscore the importance of considering gender-specific safety issues in urban transportation planning and policy-making, particularly in work-related travel.

Our study builds upon this body of work by explicitly examining work-related crashes in the context of Santiago, Chile, and evaluating the impact of a city-wide speed limit reduction on commuting safety. By focusing on this intersection of work-related mobility and urban traffic safety policies, and by analyzing the prescribed medical days of leave following crashes, collisions or falls, we contribute to a more nuanced understanding of how urban interventions can affect commuting safety and worker well-being in a Latin American context.

\section{Context and Data}

This study focuses on the Greater Santiago Metropolitan Area, the largest urban agglomeration in Chile. As of 2017, Santiago had a population of nearly 8 million inhabitants spread across an urban area of 867.75 square kilometers, comprising over 40 independent municipalities. The city's transportation system is characterized by an integrated multimodal network with a nearly flat fare structure across the metro, urban buses, and one rail service, allowing up to three transfers within a two-hour window \citep{munoz2014transantiago}.

The data for this study comes from \emph{Mutual de Seguridad de la Cámara Chilena de la Construcción} (MUSEG), one of the institutions overseen by the \emph{Superintendencia de Seguridad Social} (SUSESO) {that manages worker's insurance. Under Chilean law, this mandatory insurance covers any traffic incident (crash, collision, or fall) that occurs during a worker's commute between home and workplace, between multiple workplaces, or during work-related transportation}. 

{While police records are commonly used in traffic safety research, our insurance-based dataset offers unique advantages: 1) mandatory reporting of all non-fatal work incidents requiring medical attention, regardless of severity; 2) detailed medical outcomes through prescribed days of leave, providing a consistent measure of incident severity that captures impacts on workers' lives; and 3) consistent recording over a decade. However, the dataset has several limitations: it excludes fatal incidents (handled through a different system), covers only 31\% of insured workers with potential biases toward formal employment sectors, and like most incident databases, lacks exposure data (such as kilometers traveled) that would enable more detailed risk analysis. To address some of these limitations, we compared incident patterns with Santiago's travel survey and excluded incidents with extremely long medical leave periods ($>$2 years) to avoid outlier effects} (see Section \ref{sec:methodology} for details on data representativeness analysis).

{Our dataset spans a decade, encompassing 75.8K work-related transportation incidents in the greater Santiago Metropolitan Area from 2012 to 2021, all occurring on public roads during either commuting trips (to/from work) or work-related travel. The geographical information in our records is organized by municipalities, which are the smallest administrative subdivisions in Chile (equivalent to districts or boroughs in other countries). Each record contains:}

\begin{itemize}
    \item Date and time of occurrence
    \item Number of prescribed medical days of leave
    \item Age and gender of the involved worker
    \item Type of vehicle used at the time of the event
    \item Municipality of the worker's residence
    \item Municipality of the event
    \item Text description of the event location (usually street name or intersection)
    \item Municipality of the worker's employer
\end{itemize}

\begin{figure}[tbp]
    \centering
    \includegraphics[width=\linewidth]{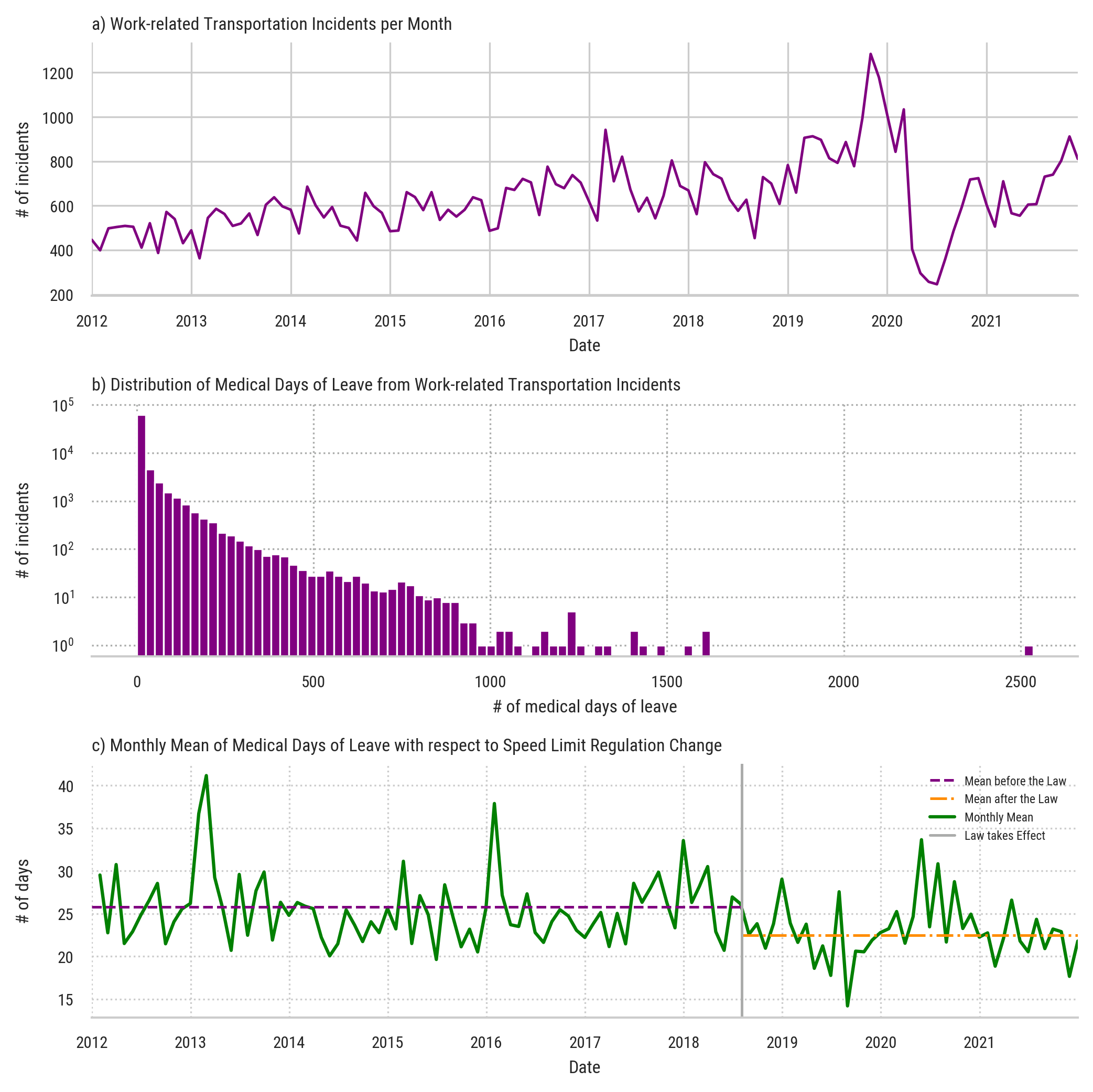}
    \caption{a) Monthly distribution of the number of crashes, collisions or falls. b) Histogram of the number of incidents concerning the number of medical days of leave. c) Mean number of prescribed medical days of leave per month, with mean value annotations before and after Chile's latest speed regulation law (August 4th, 2018).}
    \label{fig:medical_leave_distribution}
\end{figure}

{Our dataset includes three types of incidents: 1) traffic crashes between vehicles; 2) collisions between vehicles and other road users; and 3) falls during transportation (e.g., while boarding/alighting public transport or cycling).} All incidents in our analysis occurred on public roads during either commuting trips (to/from work) or work-related travel. Figure~\ref{fig:medical_leave_distribution}.a shows the monthly progression of event counts, indicating a positive trend, except for 2020, characterized by COVID-19 lockdowns.

The dataset contains non-fatal incidents only. We assess the severity of these incidents through the number of prescribed medical days of leave, {reflecting the recovery time medical professionals deemed necessary for each injured worker} (see Figure~\ref{fig:medical_leave_distribution}.b for its distribution, with a mean of 24.19 days in the whole dataset).

A significant event within our study period is the implementation of a new speed limit regulation on August 4th, 2018, which reduced the urban speed limit from 60 km/h to 50 km/h~\citep{ReduccionVelocidad}. We hypothesize that this law impacted the severity of injuries from traffic crashes, collisions or falls. The data suggests so, as the mean number of medical days of leave decreased from 25.76 before the law was in effect to 22.44 after the law was in effect (see Figure~\ref{fig:medical_leave_distribution}.c).

This rich dataset allows us to examine work-related traffic crashes, collisions or falls in Santiago from multiple perspectives, including temporal trends, spatial patterns, and demographic factors. The 2018 speed limit regulation provides a natural experiment to assess the impact of policy changes on event severity. In the following section, we detail our methodology for analyzing this data, including how we assess its representativeness and our approach to quantifying the effect of the speed limit law on event severity.

\section{Methodology}
\label{sec:methodology}

Our methodology combines statistical and spatial analysis to assess the impact of Santiago's 2018 speed limit reduction on work-related traffic incidents. We validate our dataset against the city's 2012 travel survey, analyze incident patterns, and use negative binomial regression to quantify the speed limit law's effect while controlling for transportation mode, timing, and demographics.

\subsection{Data Representativeness}

{To understand whether the work trips that resulted in transportation incidents follow similar spatial patterns as regular work trips in Santiago, we refer to the last official travel survey conducted in 2012, known as EOD2012} \citep{sectra2012informe}. This survey collected data on travel behavior, including trip origins, destinations, and modes of transportation, as well as demographic information of the resident population. {Using the worker's municipality of residence as origin and municipality of workplace as destination, we calculate the Spearman correlation coefficient between origin-destination matrices built from both MUSEG and EOD2012 data:}

\begin{equation}
r(X, Y) = \frac{\text{cov}(X_R, Y_R)}{\sigma_{X_R} \sigma_{Y_R}},
\end{equation}
where $\text{cov}$ is the covariance, and $\sigma$ is the standard deviation, applied to the rank-transformed versions of $X$ and $Y$, denoted by $X_R$ and $Y_R$. We estimate the correlation between OD matrices at the municipal level, as the EOD2012 data represents this level.

\subsection{Exploratory Analysis}

We analyze the MUSEG data concerning modal share, demographic composition, and geographical association of incidents. We focus on the daily periods defined in EOD2012:

\begin{itemize}
    \item \emph{Morning peak}: 6AM to 9AM
    \item \emph{Morning valley}: 9AM to Noon
    \item \emph{Lunch}: Noon to 2PM
    \item \emph{Afternoon Valley}: 2PM to 5:30PM
    \item \emph{Afternoon Peak}: 5:30PM to 8:30PM
    \item \emph{Nigh Valley}: 8:30PM to 11PM
    \item \emph{Night}: 11 PM to 6 AM (next day)
\end{itemize}

We also explore the demographic composition of incidents with respect to gender and age and observe if there are trends regarding severity and that demographic composition.

We use the ArcGIS Application Programming Interface to geolocate each event. We queried the API with the reported name of the street and the reported name of the incident municipality. This provides a specific coordinate for each event that can be analyzed spatially. This also allows us to determine if incidents happened in rural or urban areas. Not all incidents will be geolocated, as sometimes the reported location may be ambiguous or not present of GIS databases. 

We aggregate geolocated incidents and trip origins/destinations into a hexagonal grid with an edge length of approximately 1.4 km using the H3 system~\citep{h3geo}. {To measure spatial associations between areas and modes of transportation in incidents}, we use a log-odds ratio with an uninformative Dirichlet prior~\citep{monroe2008fightin}. {The log-odds ratio identifies areas where specific modes of transportation are over- or under-represented in incidents. Areas with few incidents but a high proportion of a specific mode (e.g., 4 out of 5 incidents being cycling-related) could suggest meaningful patterns, but raw proportions and log-odds alone do not account for sampling variability. Using an uninformative Dirichlet prior provides the needed numerical stability in areas with few or no incidents of a particular mode, while adding minimal prior information to let patterns emerge naturally from the observations. The weighted log-odds ratio is defined as follows:}
\begin{equation}
\frac{\hat{\delta}_{w}^{i-j}}{\sigma^2(\hat{\delta}_{w}^{i-j})},
\end{equation}
which is a standardized score of mode of transportation $w$ within cell $i$, with respect to all other cells, encoded as $j$. The elements $\delta$ and $\sigma$ are defined as follows:
\begin{equation}
    \begin{aligned}
        \hat{\delta}_{w}^{i-j}           & = \log(\frac{y_w^i + \alpha_w}{n^i + \alpha_0 - (y_w^i + \alpha_w)}) - \log(\frac{y_w^j + \alpha_w}{n^j + \alpha_0 - (y_w^j + \alpha_w)}) \\
        \sigma^2(\hat{\delta}_{w}^{i-j}) & \approx \frac{1}{y_w^i + \alpha_w} + \frac{1}{y_w^j + \alpha_w},
    \end{aligned}
\end{equation}
where $y_w^i$ is the observed number of incidents in mode $w$ within cell $i$, $n^i$ is the total number of incidents in cell $i$; conversely, here $j$ is interpreted as all other cells from the grid. {The value of $\alpha_w$, the prior parameter for the Dirichlet distribution, is set to 1 to be uninformative}; $\alpha_0$ represents the sum of all prior parameters for each mode $w$, $\sum_w \alpha_w$.

\subsection{Speed Limit Law Effects on Prescribed Medical Leave}

{Since all analyzed incidents occurred on public roads during commuting or work-related transportation, they were potentially affected by the speed limit change through either direct (vehicle speed) or indirect (general traffic calming) effects.}

{We use prescribed medical days of leave as a proxy measure for injury severity --- more severe injuries typically require longer recovery periods and thus more medical leave. While other severity measures exist (like Abbreviated Injury Scale or hospitalization length), our dataset provides medical leave as the primary indicator of how severely an incident impacted a worker's health and ability to return to work.}

The days of medical leave represent a highly skewed variable (see Figure \ref{fig:medical_leave_distribution}.b), making it unsuitable for typical statistical tests to compare means. Therefore, we use regression analysis, where the dependent variable is the number of prescribed medical days of leave, and the independent variables are the characteristics of the trip/event.

While count data is usually modeled with Poisson regression, our data violates the Poisson distribution's assumption that the conditional mean equals the conditional variance. We use a Negative Binomial (NB) regression model \citep{cameron1986econometric} to account for the over-dispersion in our highly skewed medical leave data. The NB model discrete probability density function is defined as:
\begin{equation}
f(y_i \mid \mathbf{x}_i) = \frac{\Gamma(y_i + \alpha)}{y_i! ~ \Gamma(\alpha)} \left(\frac{\alpha}{\alpha + u_i} \right)^\alpha \left(\frac{u_i}{\alpha + u_i} \right)^{y_i},
\end{equation}
where $y_i$ is the medical days of leave, $\mathbf{x}_i$ is the set of predictor variables, $\Gamma(\cdot)$ is the gamma function, and $\alpha$ is the dispersion parameter. The conditional mean and variance are given by:
\begin{equation}
E(y_i \mid \mathbf{x}_i) = u_i = \exp(\mathbf{x}_i^T \mathbf{\beta}); \text{Var}(y_i \mid \mathbf{x}_i) = u_i \left( 1 + \frac{1}{\alpha}u_i\right).
\end{equation}

This choice of model is coherent with previous work on analyzing behavioral and spatial patterns in Santiago \citep{graells2017effect,graells2023measuring}.

We use only geolocated incidents for the regression analysis, as these can be assigned to urban or rural areas and to specific corridors where the speed limit law does not apply, such as highways. To ensure the representativeness of this subset, we compare the distribution of medical days of leave between geolocated and non-geolocated incidents. If significant differences are found, we apply additional filtering to ensure comparability.

We focus on two specific periods for our analysis:

\begin{enumerate}
    \item Pre-law period: 2018-01-01 to 2018-07-31
    \item Post-law period: 2019-01-01 to 2019-07-31
\end{enumerate}

These periods were chosen to ensure comparability and to isolate the effect of the speed limit law. We avoid using a more extended pre-law period to prevent data imbalance. We also limit the post-law period to avoid confounding factors such as the social upheaval in Chile (from October 18th, 2019) and the COVID-19 pandemic.

To estimate the model parameters, including the $\beta$ regression coefficients and the dispersion parameter $\alpha$, we use Maximum likelihood estimation \citep{hilbe2011negative}. The independent variables in our model include:

\begin{enumerate}
    \item A boolean feature indicating whether the event occurred before or after the speed limit law implementation (our main variable of interest)
    \item Mode of transportation (e.g., car, motorcycle, cycle, pedestrian)
    \item Time of day (using the periods defined in the EOD2012 survey)
    \item Day of the week
    \item Age of the worker involved in the event
    \item Gender of the worker
    \item A boolean feature indicating whether the event occurred in a rural area or on a highway
\end{enumerate}

By including these variables, we aim to control for potential confounding factors that influence event severity independently of the speed limit change. The time-based variables (time of day, day of week) help account for variations in traffic patterns, while the demographic factors (age, gender) control for individual characteristics that might affect event outcomes. The rural/highway variable allows us to differentiate between areas directly affected by the urban speed limit change and those that were not.

We acknowledge that other unobserved factors may influence event severity during our study period. While our model cannot fully account for all possible confounders, the relatively short timeframe of our before-and-after analysis and the inclusion of multiple control variables help mitigate some of these concerns.

\section{Results}

Our analysis of work-related transportation incidents in Santiago, Chile, revealed several interesting patterns across temporal, spatial, and demographic dimensions. We present these findings, starting with the representativeness of our data, followed by exploratory analyses, and concluding with the results of our regression model examining the impact of the speed limit law.

\subsection{Data Representativeness}

To assess how well our MUSEG dataset represents overall travel patterns in Santiago, we compared it to the EOD2012 survey data. 

{According to EOD2012, there are 1.88 million workers living in Santiago, who, on an average working day, perform 13.95 million trips, of which 18.7\% are to-work trips, 2\% are for-work trips, and 46\% are to-home trips (with a fraction of these presumably mirroring work trips). We analyze the correlation between their home locations and work trips with those from MUSEG data. The Spearman correlation coefficient between the municipal home locations was $r = 0.85$ ($p <$ 0.001), and between the origin-destination matrices was $r = 0.69$ ($p <$ 0.001). These positive correlations suggests that about 85\% and 69\% respectively of the variability in the ranks of one dataset is explained by the relationship with the ranks of the other. This indicates that geographical location of worker's homes and where they commute to observed in our MUSEG dataset largely align with the general patterns in Santiago, providing confidence in the representativeness of our data.}

\begin{figure}[tbp]
    \centering
    \includegraphics[width=\linewidth]{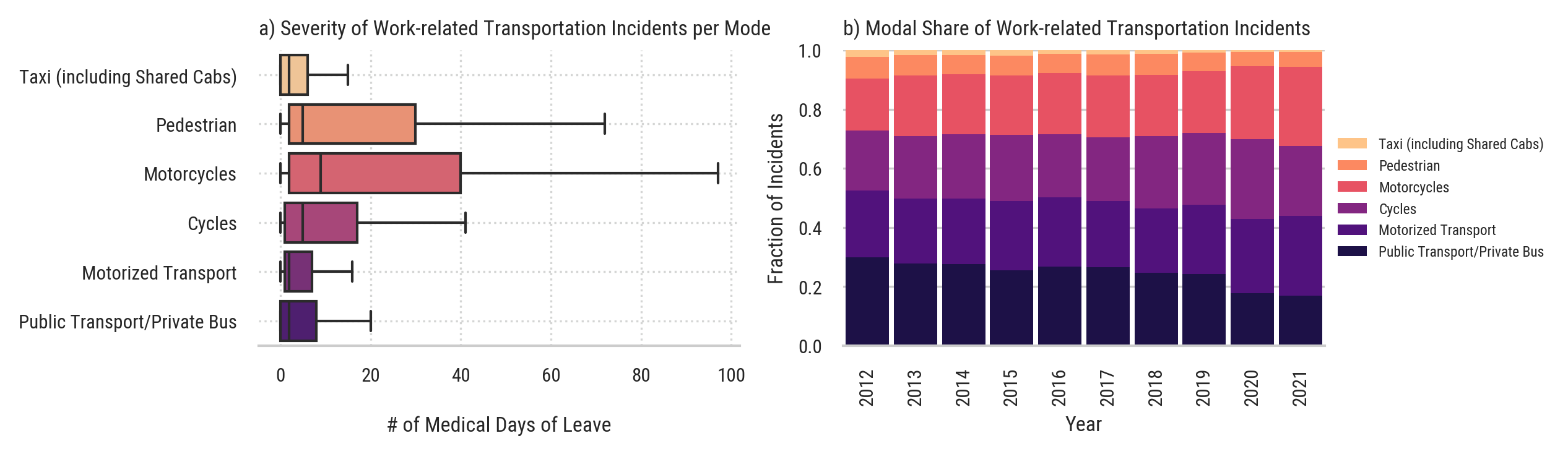}
    \caption{Mode of transportation patterns. a) A boxplot of prescribed medical days of leave in crashes, collisions, or falls with respect to mode of transportation. The midline in each box represents the median for that category. Outliers are not displayed. b) Relative distribution of mode of transportation in incidents per year.}
    \label{fig:incidents_per_year}
\end{figure}

\subsection{Mode of Transportation and Event Severity}

Our analysis of event severity by mode of transportation revealed notable differences (Figure \ref{fig:incidents_per_year}.a). Motorcycles, pedestrians and cycles were associated with the most severe incidents ({means of} 44.45, 42.08 and 23.24 prescribed medical days of leave, respectively). This pattern underscores the vulnerability of these road users.

Examining the modal share of incidents over time (Figure \ref{fig:incidents_per_year}.b), we observed a consistent decline in public transport-related incidents (from 29.84\% in 2012 to 16.99\% in 2021). Conversely, motorized and motorcycle incidents showed an increasing trend (from 22.63\% in 2012 to 26.95\% in 2021 for the former, and from 17.56\% in 2012 to 26.75\% in 2021 for the latter). These shifts align with recent findings on the evolution of modal share in Santiago, which indicate a growing preference for private modes of transportation like cars and motorcycles \citep{graells2023data}.

\subsection{Temporal Patterns}

\begin{figure}[tbp]
    \centering
    \includegraphics[width=\linewidth]{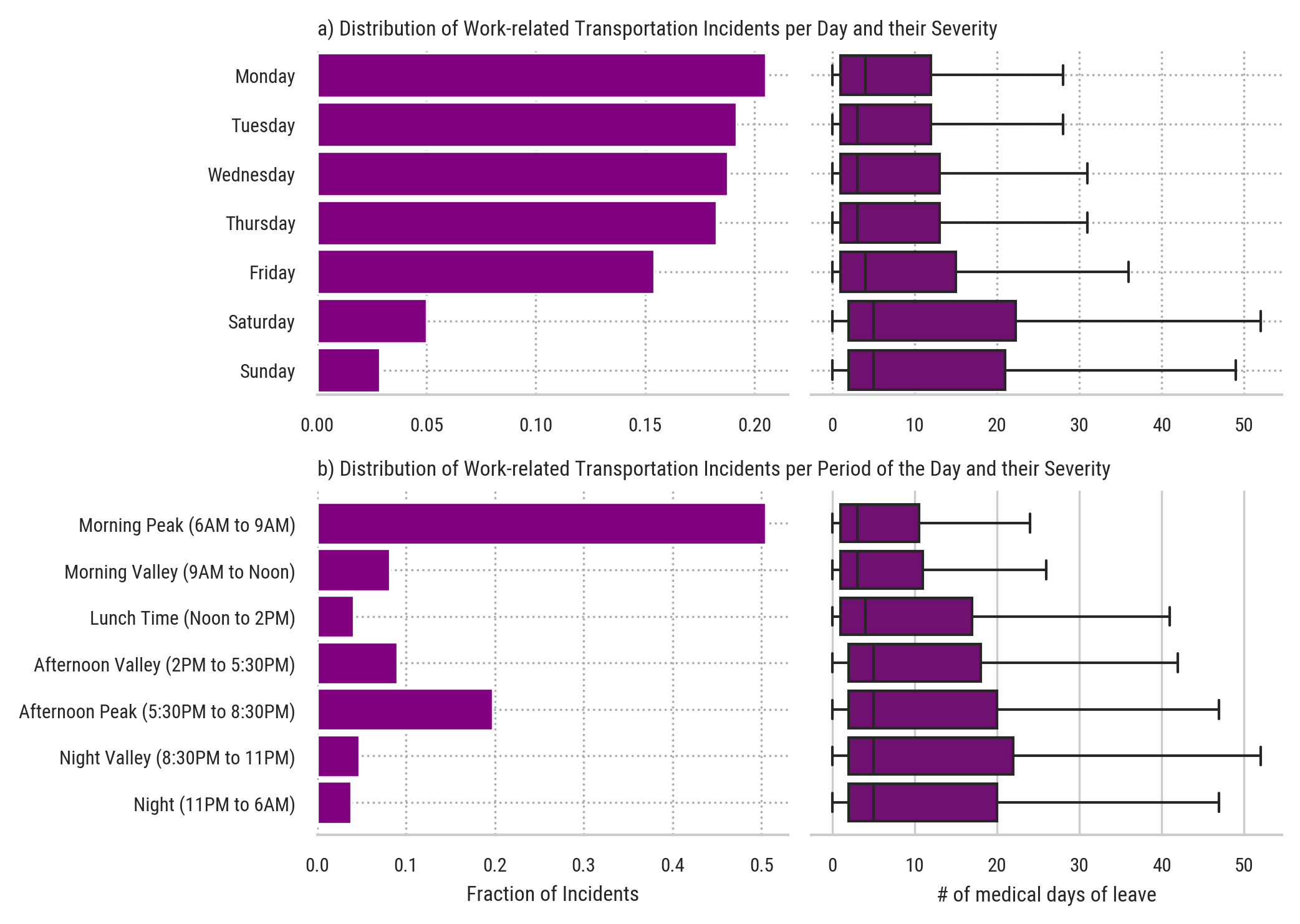}
    \caption{Incident distribution per day of the week (panel (a)) and
        period of the day as defined by the last travel survey (panel (b)).}
    \label{fig:accident_time}
\end{figure}

The temporal distribution of incidents showed clear patterns (Figure \ref{fig:accident_time}.a). Incidents were more frequent at the start of the week (76.71\% between Monday and Thursday), with a notable decrease during weekends (only 2.87\% on Sunday). This pattern likely reflects the typical workweek structure, with fewer work-related trips occurring on weekends.

Most incidents in our dataset occurred during the \emph{morning peak} hours (50.47\% of incidents, with a mean severity of 20.55 days of prescribed medical leave; see Figure \ref{fig:accident_time}.b). This concentration can be attributed to the convergence of many workers starting their day at similar times. Interestingly, while the morning peak saw the highest frequency of incidents, \emph{night} incidents (3.82\%) were associated with greater severity (32.41 days of prescribed medical leave). {This high concentration of incidents during morning peak hours likely reflects the rigid nature of work start times compared to more flexible evening departures. The greater severity of night incidents can be attributed to reduced visibility, higher vehicle speeds due to lower traffic volumes, and increased worker fatigue during night shifts. Studies of severe traffic injuries in the EU have found distinct temporal patterns, with afternoon and winter months being common times for severe pedestrian and car occupant injuries} \citep{aarts2016study}.

\begin{figure}[tbp]
    \centering
    \includegraphics[width=\linewidth]{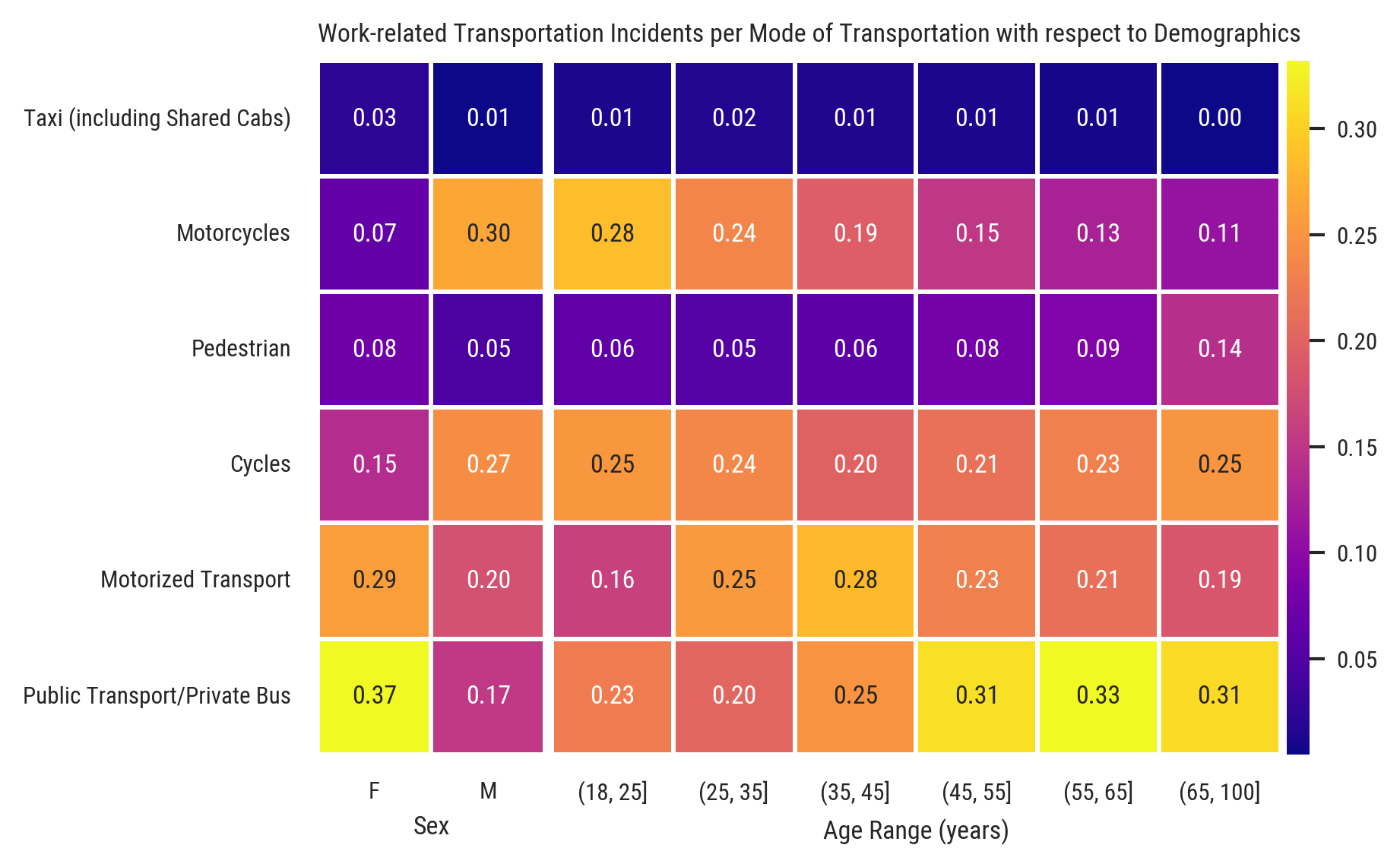}
    \caption{Fraction of incidents per gender and age ranges (column-normalized).}
    \label{fig:accident_demography}
\end{figure}

\begin{figure}[tbp]
    \centering
    \includegraphics[width=\linewidth]{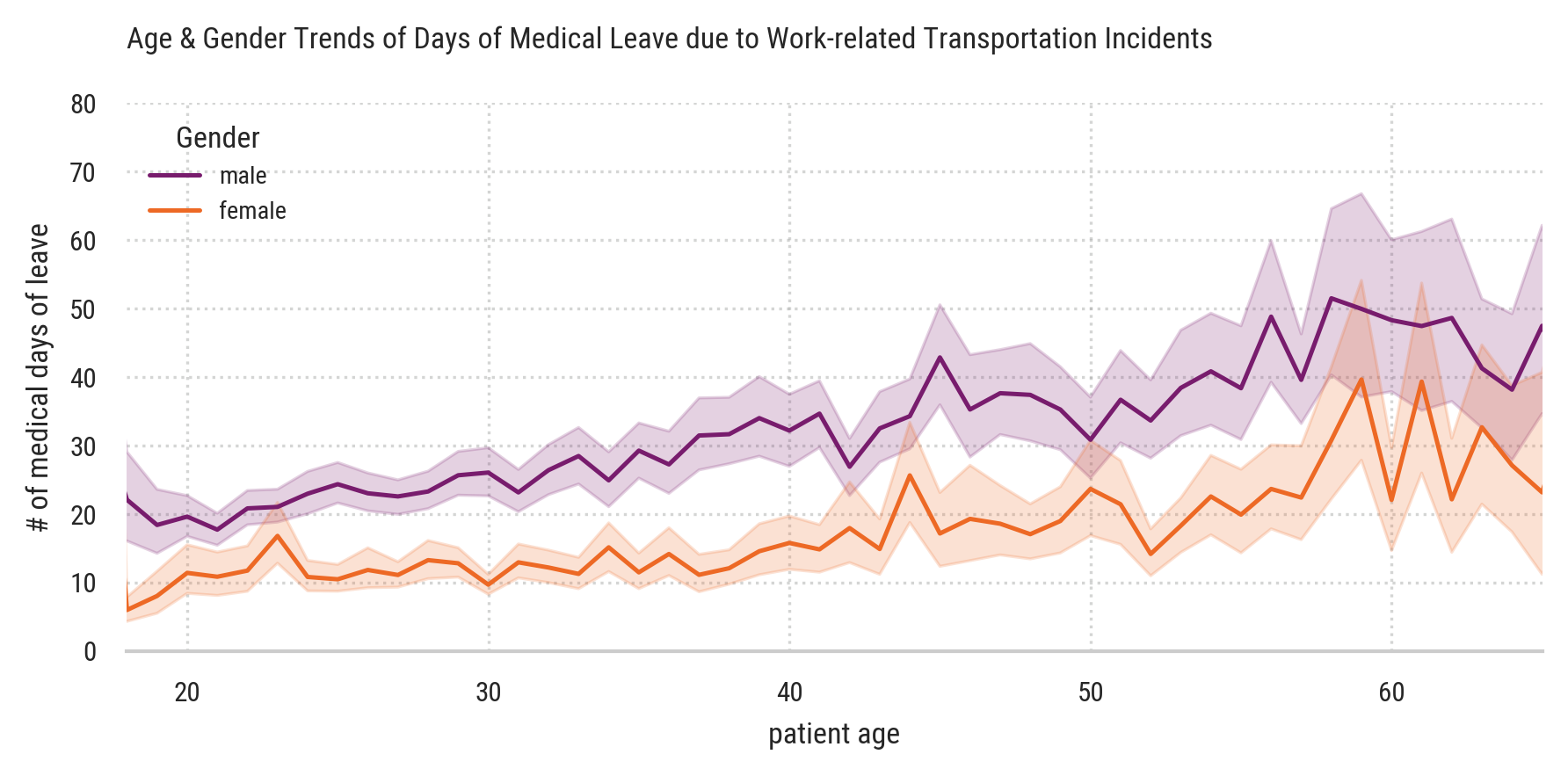}
    \caption{Age trends in the number of medical days of leave due to
        incidents, disaggregated by gender. The band surrounding each trend line
        encodes the 95\% confidence interval.}
    \label{fig:medical_leave_trends}
\end{figure}

\subsection{Demographic Patterns}
Our analysis revealed significant gender differences in incident patterns (Figure \ref{fig:accident_demography}). Women workers accounted for 37.42\% of incidents in the MUSEG data, closely mirroring their 37\% representation in the overall insured worker population \citep{mutualEstadisticas}. Similarly, the EOD2012 survey reports 36\% of commuting trips by women workers.

Gender disparities were evident in mode choice: men were involved in 87.05\% of motorcycle and 74.66\% of cycle incidents, while women were involved in 56.57\% of public transport incidents. These patterns echo gender differences observed in Santiago's latest travel survey \citep{sectra2012informe}, contrasting the worker composition ascribed to MUSEG, where 63\% is men and 37\% is women.

{The analysis of incident distribution across age groups} (Figure \ref{fig:accident_demography}) {reveals distinct modal patterns. For motorcycle incidents, the proportion decreases with age, from 0.28 in the 18--25 age group to 0.13 in the 55--65 group. Conversely, public transportation incidents increase with age, from 0.23 to 0.33 in the same age ranges. Similar age-related increases are observed for pedestrian (0.06 to 0.09) and motorized incidents (0.16 to 0.21, peaking at 0.28 in the 35--45 age group). These shifts suggest age-dependent changes in transportation mode choice, though a detailed analysis of this relationship is beyond the scope of this study.}

{Regarding incident severity, we found that men workers tended to experience more severe incidents} (Figure \ref{fig:medical_leave_trends}). {Similar gender differences in medical leave duration were found by} \citet{bermudez2018discrete}. {Additionally, incident severity increased with age for both genders. The detailed quantification of these demographic effects is presented in the regression analysis section. However, without exposure data (e.g., kilometers traveled by gender), we cannot make conclusions about relative risk levels across demographic groups. The patterns described reflect incident characteristics rather than differential risk rates.}

\subsection{Spatial Patterns}

\begin{figure*}[tbp]
    \centering
    \includegraphics[width=0.99\linewidth]{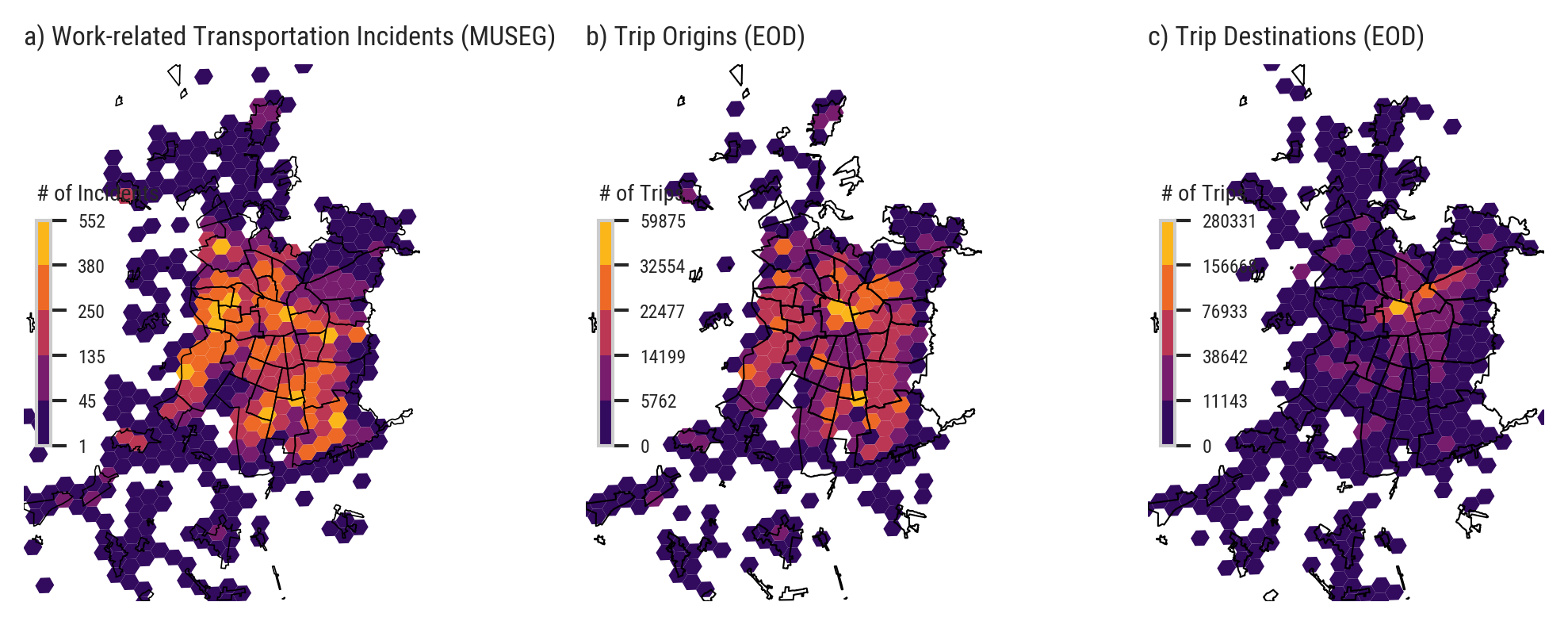}
    \caption{Spatial distributions of incidents in the MUSEG dataset (a), and commuting trip origins (b) and destinations (c) according to the EOD2012 dataset.
        Brighter colors encode a greater amount of incidents or trips. Black
        lines depict administrative boundaries for urban areas, having into account municipal boundaries.}
    \label{fig:accident_eod}
\end{figure*}

\begin{figure*}[tbp]
    \centering
    \includegraphics[width=\linewidth]{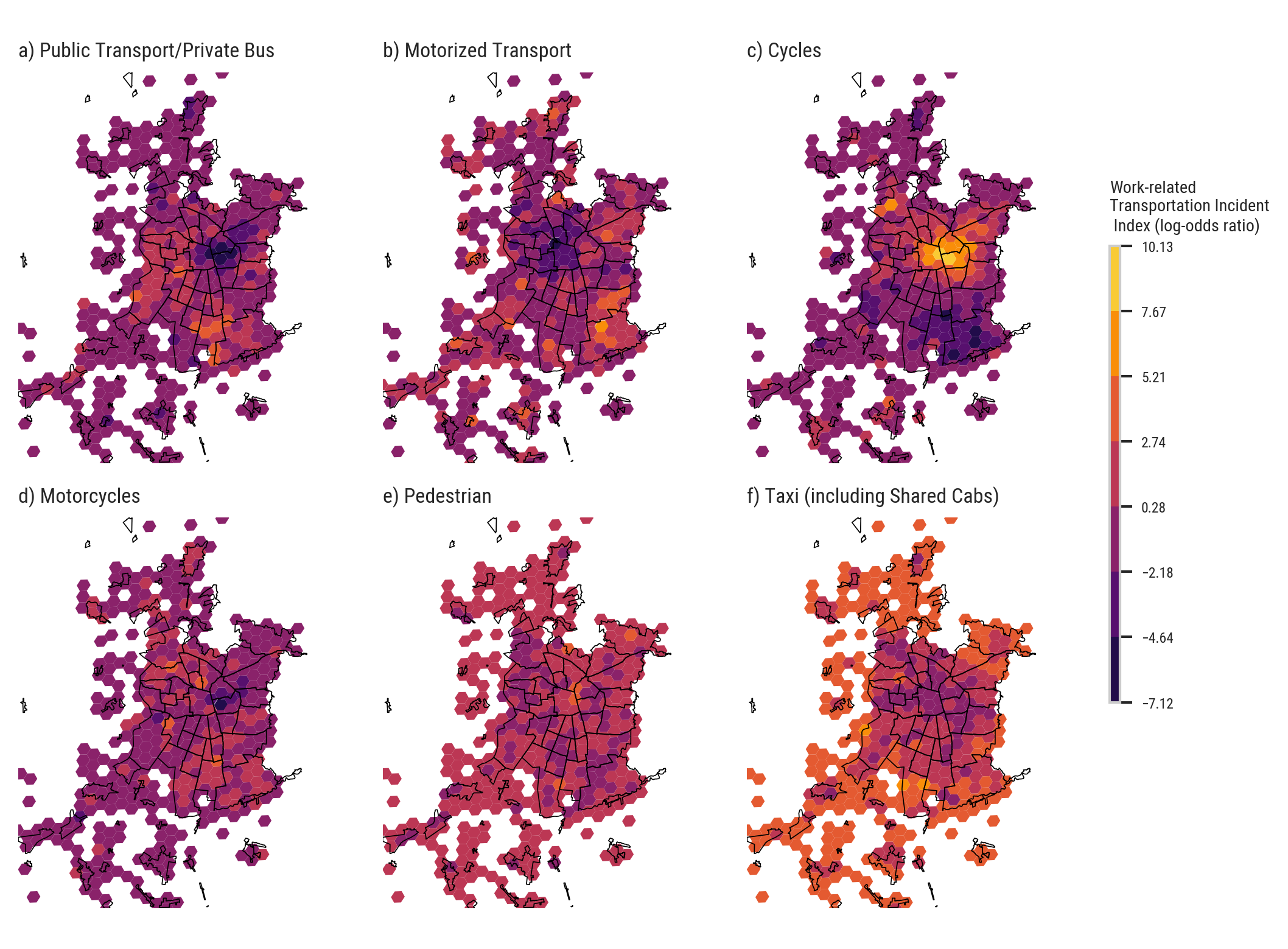}
    \caption{Spatial associations with mode of transportation in each
        incident with cells from the H3 grid system. The association is defined
        as a log-odds ratio with an uninformative Dirichlet prior
        \citep{monroe2008fightin}. Each mode of transportation has its own map:
        a) Public Transportation and Private Buses; b) Motorized Transport
        (private cars and other vehicles with four wheels); c) Cycles (including
        bikes and scooters); d) Motorcycles; e) Pedestrian; and f) Taxi,
        including shared cabs, which are used predominantly in the south of the
        city. Positive log-odds ratios, depicted in shades of orange, signify
        areas where incidents in the respective mode are more prevalent than
        expected. Conversely, negative ratios, depicted in shades of purple,
        indicate fewer incidents than anticipated.}
    \label{fig:accident_geography}
\end{figure*}

We geolocated 33.4K (44.05\%) incidents in the Santiago area from 2012 to 2021. Our spatial analysis of incidents (Figure \ref{fig:accident_eod}.a) revealed a distribution that differed from both the origins (Figure \ref{fig:accident_eod}.b) and destinations (Figure \ref{fig:accident_eod}.c) of commuting trips reported in the EOD2012 survey. This discrepancy is expected, as incidents can occur at any point during a commute, not just at trip endpoints.

When examining incident patterns by mode of transportation (Figure \ref{fig:accident_geography}), we found interesting spatial associations. Cycle incidents clustered in zones with well-developed cycleways and significant poles of attraction for work-related trips. In contrast, motorized vehicles (e.g., cars) exhibited a stronger association with the city's peripheral sectors. These areas serve either as trip generators or intermediate pass-through zones, potentially leading to higher vehicle speeds compared to other city areas. {Using log-odds ratios, we found cycle incidents were more strongly associated with zones that have well-developed cycleways and significant poles of attraction for work-related trips, though our analysis does not assess the relationship between infrastructure quality and safety.} {While this analysis shows relative concentrations of incidents, it cannot determine actual risk rates without exposure data (trips or kilometers traveled by mode in each zone).}

\subsection{Speed Limit Law Effects}

\begin{figure}[tbp]
    \centering
    \includegraphics[width=\linewidth]{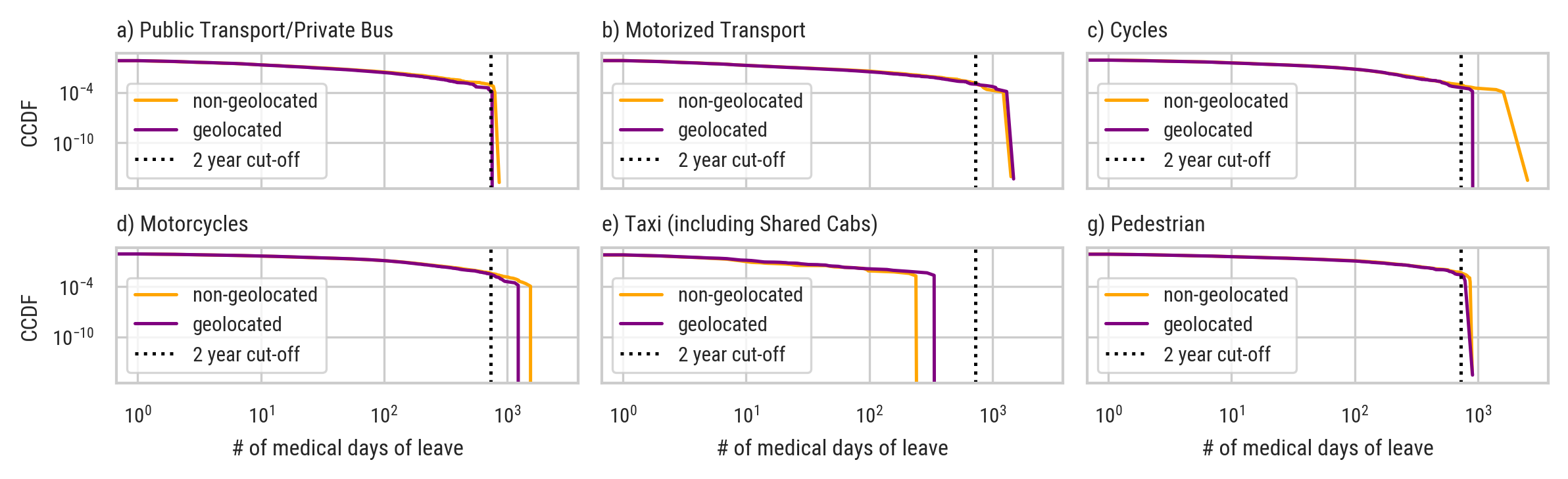}
    \caption{Complimentary cumulative density function of the number of medical days of leave, per mode of transportation. Each image compares the distribution of geolocated and non-geolocated incidents in the area of study.}
    \label{fig:mode_ccdf}
\end{figure}

For our regression analysis, only 1.06\% of the geolocated incidents occurred in rural areas, and 1.07\% were located near highways, areas not directly impacted by the urban speed limit law.

Before proceeding with the regression, we compared the distribution of medical days of leave between geolocated and non-geolocated incidents (Figure \ref{fig:mode_ccdf}). The distributions were similar for incidents resulting in up to two years of medical leave. Consequently, we excluded the small fraction of incidents (0.001\%) with longer leave periods from our analysis. We also excluded taxi trips (1.32\% of the dataset) due to their distinct distribution pattern. In total, we discarded 1.24\% of observations according to this criteria.

\begin{table}[tbp]
    \caption{Summary table of regression results in the Negative Binomial Regression.}
    \label{table:nb_regression}
    \centering
    \footnotesize
    \begin{tabular}{@{}|l|l|l|l|@{}}
        \toprule
        No. Observations: & 4620      & Df Residuals:   & 4602    \\
        \midrule
        Method:           & MLE       & Df Model:       & 17      \\
        \midrule
        Pseudo R-squ.:    & 0.01831   & Log-Likelihood: & -15341  \\
        \midrule
        converged:        & True      & LL-Null:        & -15627  \\
        \midrule
        Covariance Type:  & nonrobust & LLR $p$-value:    & $<$ 0.001 \\
        \bottomrule
    \end{tabular}

    \raggedleft
    \begin{tabular}{@{}p{\dimexpr 0.3\linewidth-2\tabcolsep}lllllll@{}}
        \toprule
                                                      & Coeff.  & S.E.  & $z$    & $p$       & \textbf{[0.025} & \textbf{0.975]}   \\ \midrule
        Intercept                                     & 1.4738  &        0.145     &    10.179  &         $<$ 0.001        &        1.190    &        1.758  \\
        \midrule Women (w.r.t. Men)                 & -0.3588  &        0.063     &    -5.718  &         $<$ 0.001        &       -0.482    &       -0.236 \\
        Age (in years)                                & 0.0269  &        0.002     &    11.083  &         $<$ 0.001        &        0.022    &        0.032 \\
        \midrule Bikes \& Scooters (w.r.t. Motorized) & 0.6193  &        0.086     &     7.177  &         $<$ 0.001        &        0.450    &        0.788  \\
        Pedestrian (w.r.t. Motorized)                 & 0.8366  &        0.127     &     6.568  &         $<$ 0.001        &        0.587    &        1.086  \\
        Motorcycle (w.r.t. Motorized)                 & 1.2772  &        0.089     &    14.427  &         $<$ 0.001        &        1.104    &        1.451  \\
        Public/Massive Transport (w.r.t. Motorized)   & -0.0174  &        0.081     &    -0.215  &         0.830        &       -0.176    &        0.141  \\
        \midrule Morning Peak (w.r.t. Morning Valley) & -0.0921  &        0.098     &    -0.939  &         0.348        &       -0.284    &        0.100  \\
        Lunch Time (w.r.t. Morning Valley)            & 0.1192  &        0.151     &     0.790  &         0.429        &       -0.177    &        0.415  \\
        Afternoon Valley (w.r.t. Morning Valley)      & 0.5169  &        0.121     &     4.282  &         $<$ 0.001        &        0.280    &        0.754  \\
        Afternoon Peak (w.r.t. Morning Valley)        & 0.2618  &        0.109     &     2.391  &         0.017        &        0.047    &        0.476  \\
        Night Valley (w.r.t. Morning Valley)          & 0.2629  &        0.183     &     1.436  &         0.151        &       -0.096    &        0.622  \\
        Night (w.r.t. Morning Valley)                 & 0.5314  &        0.159     &     3.350  &         0.001        &        0.221    &        0.842  \\
            \midrule
        Friday (w.r.t. Monday-Thursday)               & -0.0751  &        0.082     &    -0.917  &         0.359        &       -0.236    &        0.085  \\
        Saturday (w.r.t. Monday-Thursday)             & 0.2153  &        0.123     &     1.754  &         0.079        &       -0.025    &        0.456  \\
        Sunday (w.r.t. Monday-Thursday)               & 0.1678  &        0.164     &     1.025  &         0.306        &       -0.153    &        0.489  \\
        \midrule Rural Area or Highway                & 0.4302  &        0.203     &     2.118  &         0.034        &        0.032    &        0.828  \\
        Speed Limit Law                               & -0.2021  &        0.057     &    -3.534  &         $<$ 0.001        &       -0.314    &       -0.090 \\
        \midrule $\alpha$                             & 3.5531  &        0.076     &    47.048  &         $<$ 0.001        &        3.405    &        3.701  \\ \bottomrule
    \end{tabular}
\end{table}

Our final regression dataset comprised 4,620 incidents from the first seven months of 2018 and 2019. We performed the regression analysis using Python with the \emph{statsmodels} library \citep{seabold2010statsmodels}. Table \ref{table:nb_regression} presents the results of our Negative Binomial regression model.

\begin{figure}[tbp]
    \centering
    \includegraphics[width=\linewidth]{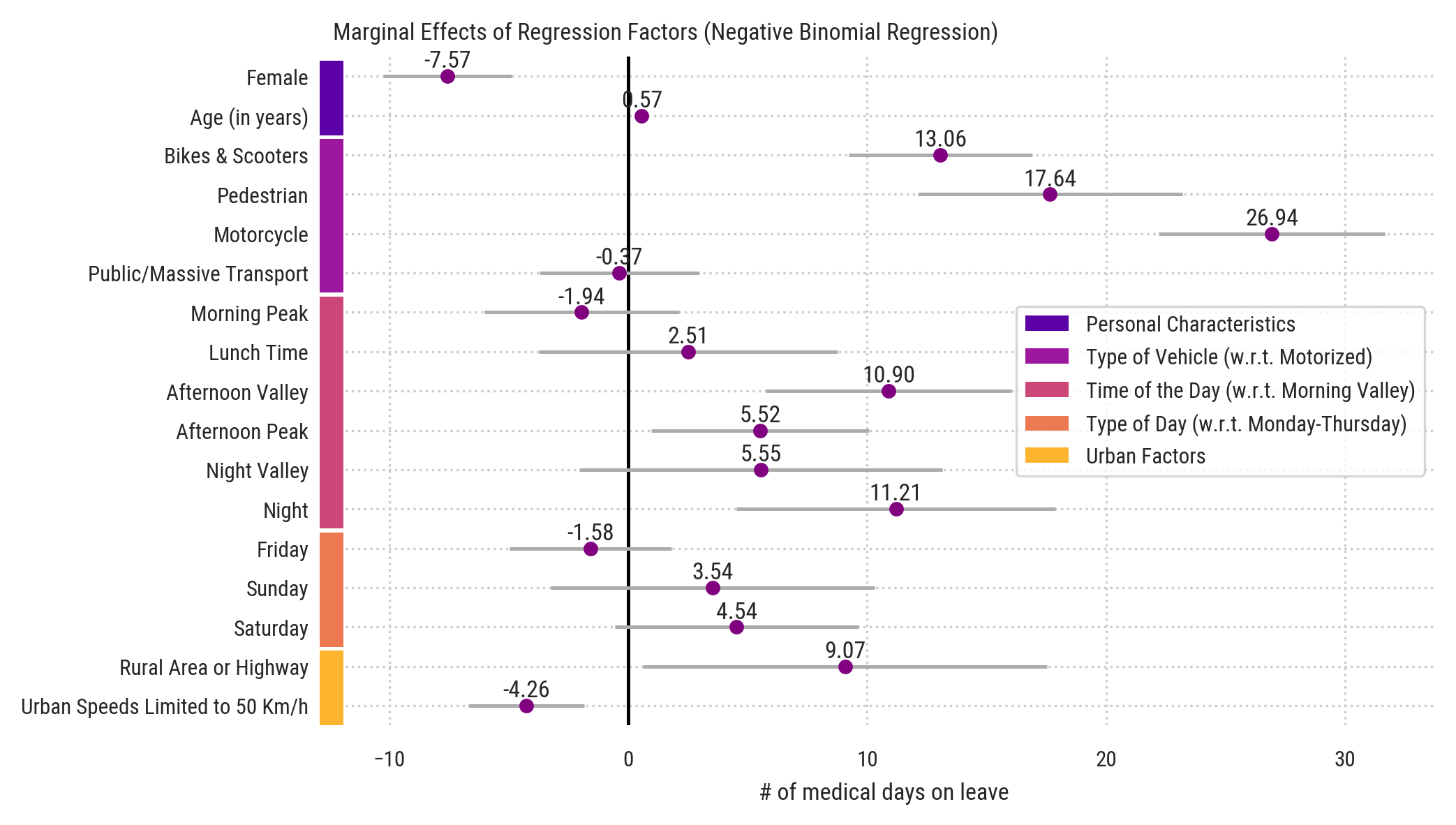}
    \caption{Marginal effects at the mean of regression factors. Factors are grouped according to a high-level category. Their marginal effect is represented by a dot with error bars that represent the corresponding 95\% confidence interval.}
    \label{fig:marginal_effects}
\end{figure}

We calculated the marginal effects at the mean for each variable to ease interpretation of the results (Figure \ref{fig:marginal_effects}). The marginal effect represents the change in the expected number of prescribed medical days of leave for a one-unit change in the independent variable, holding all other variables constant at their mean values. For categorical variables, it represents the difference in expected days of leave when the variable is true versus false, or when compared to the reference category. A positive marginal effect implies more severe injuries, as it is associated with more prescribed days of medical leave.

The model revealed several significant factors influencing incident severity:

\begin{enumerate}
\item Mode of Transportation: Compared to motorized vehicles, cycles (marginal effect of 13.06 additional days of leave), motorcycles (26.94 days), and pedestrian trips (17.64 days) were associated with more severe injuries. This aligns with our exploratory findings and highlights the vulnerability of these users, consistent with previous studies on risk \citep{miner2024car, mangones2024safety}.
\item Demographics: Women workers tended to have less severe injuries (7.57 fewer days of leave on average). This gender difference in injury severity has been observed in other studies \citep{dejoy1992examination, gonzalez2021traffic, havet2021gender}. Age was also a significant factor, with each year associated with 0.57 additional days of leave, reflecting the increased vulnerability of older workers to severe injuries.
\item Time of Day: incidents during the \emph{afternoon valley} (10.90 additional days), \emph{afternoon peak} (5.52 days), and \emph{night} (11.21 days) periods were associated with more severe injuries compared to the morning valley period, a result consistent with differences in driving at different periods of the day \citep{mohammadi2009pattern} and in severity of incidents \citep{aarts2016study}.
\item Location: incidents in rural areas or on highways were more severe (9.07 additional days of leave), consistent with other studies \citep{cabrera2020uncovering}.
\item Speed Limit Law: Importantly, our analysis showed that the speed limit law was associated with a decrease of 4.26 days in prescribed medical leave for incidents in urban areas. This suggests that the law had a positive impact on reducing injury severity in work-related transportation incidents.
\end{enumerate}

These results, visualized in Figure \ref{fig:marginal_effects}, provide valuable insights into the factors influencing injury severity in Santiago and the potential benefits of speed management policies in urban areas. The marginal effects allow us to quantify the relative impact of each factor on injury severity, providing a clear picture of which elements have the strongest influence on the length of medical leave following a work-related transportation incident.

\section{Conclusions}

Our analysis of work-related transportation incidents in Santiago, Chile, over a decade, revealed significant patterns in incident occurrence and severity across temporal, spatial, and demographic dimensions. These findings have important implications for urban planning, transportation policy, and workplace safety initiatives.

A key finding of our study is the impact of the 2018 urban speed limit reduction. This policy change was associated with a decrease in injury severity, with incidents requiring an average of 4.26 fewer days of medical leave. This result provides empirical support for the effectiveness of speed management policies in improving road safety \citep{elvik2012speed, cleland2020effects, milton2021use}. 
The observed increase in car usage in Santiago during the study period aligns with trends reported by \citet{graells2023data}. 
{This finding is particularly significant as it provides evidence of speed limit effectiveness in a Latin American context, where such evaluations are scarce.}
{While our results show an overall reduction in injury severity after the speed limit change, we cannot fully determine if the risk per user decreased for each mode of transportation. Changes in mode share during our study period mean that apparent reductions in incidents for certain modes could mask increased per-user risk if there were fewer total users of that mode. This is particularly relevant for vulnerable road users like pedestrians and cyclists.}

The study identified motorcycles, cycles, and pedestrian trips as modes associated with more severe injuries compared to motorized vehicles. This vulnerability of certain users highlights the need for targeted safety measures \citep{miner2024car, mangones2024safety}. We observed gender differences in incident severity, with women workers experiencing less severe injuries on average. Older workers were more likely to suffer more severe injuries \citep{dejoy1992examination, gonzalez2021traffic, elrud2019sickness}. These demographic factors suggest potential benefits from age and gender-specific safety interventions.

Temporal patterns emerged, with incidents during afternoon and night periods associated with more severe injuries, aligning with previous studies on severity factors \citep{aarts2016study}. Spatially, incidents in rural areas or on highways were more severe \citep{cabrera2020uncovering}, indicating a need for tailored safety strategies for different road environments.

{This research demonstrates the effectiveness of speed management policies in improving urban mobility and safety.}
\citet{Guzman2020Confronting} developed Sustainable Mobility Plans for organizations in Bogot\'a, highlighting the need for tailored strategies to address congestion and enhance safety in urban areas. The results have practical implications for policymakers and urban planners. The positive impact of the speed limit reduction supports expanding such policies in urban areas. The higher injury severity for certain road users calls for infrastructure improvements and safety measures to protect vulnerable groups. {While we observed differences in incident patterns across demographic groups, additional research incorporating exposure data (e.g., from contemporary travel surveys) would be needed to understand differential risks and design targeted interventions.}

Integrating road safety into broader urban design strategies aligns with `safe system' approaches to urban mobility \citep{wegman2012make}. A comprehensive approach to urban mobility safety should combine infrastructure improvements with education, enforcement, and policy measures \citep{banister2008sustainable}. Future urban planning should consider these complex interactions to create safer urban environments.

Our study has limitations. The data comes from a single insurance provider, which may only partially represent some work-related incidents in Santiago. {Our dataset lacks exposure data (such as kilometers traveled by different demographic groups), which limits our analysis of differential risk patterns}. Our analysis does not include fatal incidents, which may have different patterns and risk factors. {Furthermore, our dataset did not capture information about other vehicles involved in multi-vehicle incidents, preventing analysis of how different vehicle types contribute to injury severity. Studies have shown that larger vehicles pose disproportionate risks to other road users} \citep{aldred2021does}, {suggesting this is a crucial aspect for understanding traffic safety}. Future research could incorporate data from multiple sources, include fatal incidents{, and examine vehicle interaction patterns} for a more comprehensive analysis. Our analysis relied on geolocated incidents, although a relevant fraction of incidents were not geolocated. We analyzed a specific subset of geolocated incidents to ensure experimental validity; however, future work could deepen the analysis by exploring individual {administrative regions}. In this case, all data could be used. {Additionally, our spatial analysis cannot account for exposure levels by mode, preventing assessment of actual risk rates in different areas.}

In conclusion, our findings highlight the complex relationship between urban design, transportation policy, and road safety. The demonstrated impact of the speed limit reduction on injury severity illustrates how policy interventions can improve safety outcomes for all road users, even in the context of increasing car usage. Integrating these insights into urban planning and transportation policy can contribute to safer, more sustainable urban mobility.

\bibliographystyle{elsarticle-harv}
\bibliography{cas-refs}

\if@endfloat\clearpage\processdelayedfloats\clearpage\fi






\end{document}